\documentclass[
final            
  ]
  {aipproc}
\usepackage{amssymb}
\usepackage{multicol}
\usepackage{graphicx}
\layoutstyle{6x9}

\begin{document}

\title{The Pati-Salam Supersymmetric SM (PSSSM)}

\classification{11.30.Qc,12.10.Dm,12.60.Jv,14.80.Ly}

\keywords{SUSY GUTs, chiral exotics, multi-dimensional parameter spaces}

\author{Felix Braam}{
  address={University of Freiburg, Institute of Physics,
  Hermann-Herder-Str. 3, 79104 Freiburg, Germany}
}

\author{J\"urgen Reuter}{
  address={University of Freiburg, Institute of Physics,
  Hermann-Herder-Str. 3, 79104 Freiburg, Germany}
}
\author{Daniel Wiesler}{
  address={University of Freiburg, Institute of Physics,
  Hermann-Herder-Str. 3, 79104 Freiburg, Germany}
}

\begin{abstract}
Grand Unified Theories based on an $E_6$ gauge group embed all Higgs-
and matter fields within one $\mathbf{27}$ dimensional fundamental
representation. In addition each $\mathbf{27}$ contains a right handed
neutrino, an NMSSM-like standard model singlet and a pair of exotic
particles, that couple quarks to leptons. The extended particle
content spoils simple gauge unification as in the MSSM. By embedding
the SM into a Pati-Salam gauge group above $10^{16}$ GeV, one obtains
a unification below the Planck-scale. We present a Markov chain
Monte-Carlo algorithm to systematically scan the high dimensional
parameter space and calculate low-energy spectra, as well as first
results of our studies of the LHC phenomenology of leptoquarks and
leptoquarkinos with the event generator WHIZARD. 

\end{abstract}

\maketitle


\section{$E_6$ GUT with Intermediate Pati-Salam symmetry}

Grand unified theories (GUT) in the framework of supersymmetry have been 
investigated for decades. 
Under the assumption of a minimal supersymmetric extension (MSSM)
of the standard model (SM), the SM gauge couplings unify at approximately
$10^{16}$ GeV. At higher energies, the known interactions might be described 
by a single simple gauge group. However, the choice of the unified gauge group
is not unique and potentially interferes with the low-energy theory, as shall 
be illustrated in the following.

Throughout this note, we assume a GUT with the rank 6 exceptional Lie-group 
$E_6$ as gauge group. Decomposing the $\mathbf{27}$ dimensional fundamental 
representation under its SM subgroup, one observes that the $\mathbf{27}$ 
contains, in addition to one full 
generation of SM quarks and leptons,
a right-handed neutrino as well as a full MSSM Higgs-sector,
augmented by a SM singlet $S$ as in the NMSSM. 
Furthermore there is a pair of exotic particles
$D,\,D^c$ that may couple quarks to leptons and shall be denoted as 
leptoquarks.\footnote{These particles could also have diquark couplings -  
only one should be allowed, in order to prevent the proton to decay. In this 
work, we focus on the leptoquark case. }

Without the introduction of a further mechanism to render the exotic particle 
heavy at the $E_6$-breaking scale, the full particle content remains in the 
spectrum down to the TeV-scale \cite{Kilian:2006hh}. 

The extra matter changes the renormalization group equations and hence the 
running of the couplings: Simple unification of all three SM couplings as 
in the MSSM is spoiled.

As proposed in \cite{Kilian:2006hh} and furtherly investigated in
\cite{Howl:2007hq}, unification to a simple gauge group can be 
achieved by introducing an intermediate semi-simple gauge group,
namely a  Pati-Salam symmetry \cite{Pati:1974yy}:
\begin{equation}
SU(4)\times SU(2)_L\times SU(2)_R[\times U(1)_\chi]
\label{PS-group}
\end{equation} 
In this regime there are only two independent couplings left to intersect, 
as $SU(2)_L$ and $SU(2)_R$ have identical particle content.

A suitable breaking of the intermediate gauge symmetry (\ref{PS-group}) to the 
SM gauge group is achieved by 
introducing some Higgs fields $H^\prime,\,\bar{H}^\prime$ that
transform as $\mathbf{27}, \overline{\mathbf{27}}$ under $E_6$ and aquire a
Vev in the direction of the right-handed neutrino.
This generates Majorana masses for the right-handed neutrinos, arising from the
effective operator $(\mathbf{27\,78\,}\overline{\mathbf{27}})^2$
which trigger a see-saw mechanism to account for naturally small neutrino masses.      
The SM hypercharge group becomes a linear combination of the Cartan generators
$T^{15}\,\propto\, (B-L)/2$ of $SU(4)$ and $T^3_R$: $Y\,=\,(B-L)/2+T^3_R$.
The $SU(3)_c$ group is just the evident subgroup of $SU(4)$. Finally, 
$U(1)^\prime$ corresponds to the unbroken linear combination of 
$(B-L)/2,\,T^3_R$ and $Q_\chi$ orthogonal to $Y$ at the scale where the 
PS-group is broken down to the SM.\footnote{The requirement of 
the orthogonality to $Y$ yields non-rational charges $Q^\prime$. Furthermore, 
the RGE evolution of the two $U(1)$ couplings introduces a non-trivial mixing 
among these groups, which can be taken into account by redefining $Q^\prime$. }
The full unification scenario is shown in Figure 1.

\section{Low-energy spectra and LHC Phenomenology}

The superpotential is obtained from decomposing an $E_6$-symmetric 
superpotential $W_{E_6}\propto\mathbf{27}^3$, omitting the terms involving the 
right handed neutrinos as well as the baryon-number violating interactions of
the exotic particles $D,\,D^c$:
\begin{equation}
 W = W_{\mbox{\tiny{MSSM}}} +  \mathbf{Y}^{SH} S\,H^uH^d     
                               +  \mathbf{Y}^{SD} S\,D\,D^c  
  + \mathbf{Y}^{D_1} D\, u^c e^c  + \mathbf{Y}^{D_2} D\, d^c \nu^c
  + \mathbf{Y}^{D^c} D^c L \,Q. 
\end{equation}
A priori, all Yukawa couplings are rank three tensors in family space. However,
in order to avoid flavor changing neutral currents, we introduce a so-called 
H-parity \cite{Griest:1990vh}, that only allows one Higgs generation to acquire a 
Vev and couple to matter. This parity renders the lightest ``un-Higgs'' stable 
and hence introduces a possible additional type of dark matter.

\begin{figure}
\begin{minipage}{.95\linewidth}
  \raisebox{-2cm}{\mbox{\includegraphics[width=.40\linewidth]{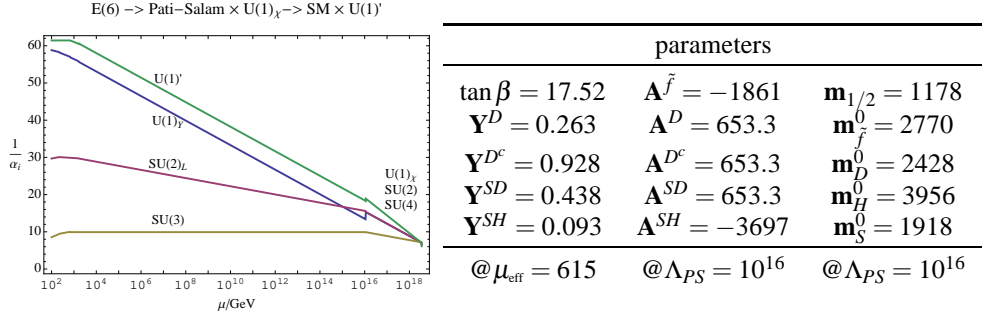}}}
 \begin{footnotesize}
 \begin{tabular}[width=.40\linewidth]{ccc}
 \hline& parameters  \\
 \hline
       $  \tan{\beta}=17.52$      & $\mathbf{A}^{\tilde{f}}=-1861$
      & $\mathbf{m}_{1/2}=1178 $
      \\
        $ \mathbf{Y}^{D}=0.263$    & $ \mathbf{A}^{D}=653.3$
      & $ \mathbf{m}^0_{\tilde{f}}=2770$
      \\
        $ \mathbf{Y}^{D^c}=0.928$  & $ \mathbf{A}^{D^c}=653.3$
      & $ \mathbf{m}^0_{D}=2428$
      \\
        $ \mathbf{Y}^{SD}=0.438$   & $ \mathbf{A}^{SD}=653.3$
      & $ \mathbf{m}^0_{H}=3956$
      \\
        $ \mathbf{Y}^{SH}=0.093$   & $ \mathbf{A}^{SH}=-3697$
      & $ \mathbf{m}^0_{S}=1918$ 
      \\ \hline
       $@ \mu_{\mbox{\tiny{eff}}}  = 615$
         & $@ \Lambda_{PS}=10^{16}$ & $@ \Lambda_{PS}=10^{16}$  
 \end{tabular}
 \end{footnotesize}
\end{minipage}
 \caption{\label{tab_params}
 Left: The SM gauge groups are unified at $\sim10^{16}$ GeV to the
 Pati-Salam group (\ref{PS-group}). Full $E_6$ unification is realized
 at $\sim10^{18}$ GeV. 
 Right: Sample input parameters and the scales they are defined at.
}
\end{figure}

As in most SUSY models (in fact all BSM models without a theory about
flavor), the model contains a huge number of free parameters in  
the SSB sector. Assuming the trilinear soft couplings to be
proportional to the corresponding Yukawas, the latter to be diagonal
and neglecting complex phases, we arrive  at a set of 16 parameters
(see Table \ref{tab_params}). Three of those are be fixed by the minimization of 
the Higgs-potential. From this parameter set, a code generates a spectrum by 
solving the RGEs, minimizing the Higgs potential and iterating in the 
dependent parameters. Throughout a run, it checks for perturbativity of the 
theory up to the unification scale and negative scalar mass-squares. 
In the end, the spectrum is compared to the current direct experimental bounds.

This algorithm is integrated into a Monte-Carlo Markov-Chain (MCMC), in order 
to efficiently scan the high-dimensional parameter space. Using this method, we
hope to find a somewhat representative set of scenarios to characterize the 
phenomenology of this model.

General phenomenological features of this model are: a more or less
hierarchical structure of leptoquarks in the range from 500 GeV to
several TeV as their masses $m_D\propto \langle S_3\rangle$ are
proportional to the singlet vev. This has to be big enough (which
comes more or less for free on the other hand, driven by a possibly
large leptoquark singlet Yukawa coupling) in order to have a $Z'$
above the current Tevatron limits and electroweak precision data
limits. For the very same reason, the so-called un-Higgses (two of the
three CP-even), the charged as well as the CP-odd Higgses are rather
heavy, i.e. $M_H \gtrsim 1$ TeV. In contrast, the lightest Higgs boson
runs quite often in the NMSSM parameter channel where it can be as
light as $90$ GeV in some regions of the parameter space. As in the
NMSSM, the decays via the lightest pseudoscalars into four $b$ jets
which already excluded discovery at LEP, render a discovery at LHC very
intricate. Another very distinct feature is a rather light gluino not
much heavier than weak inos, because the QCD beta function vanishes at
first order.

\begin{figure}
\includegraphics[width=.45\linewidth]{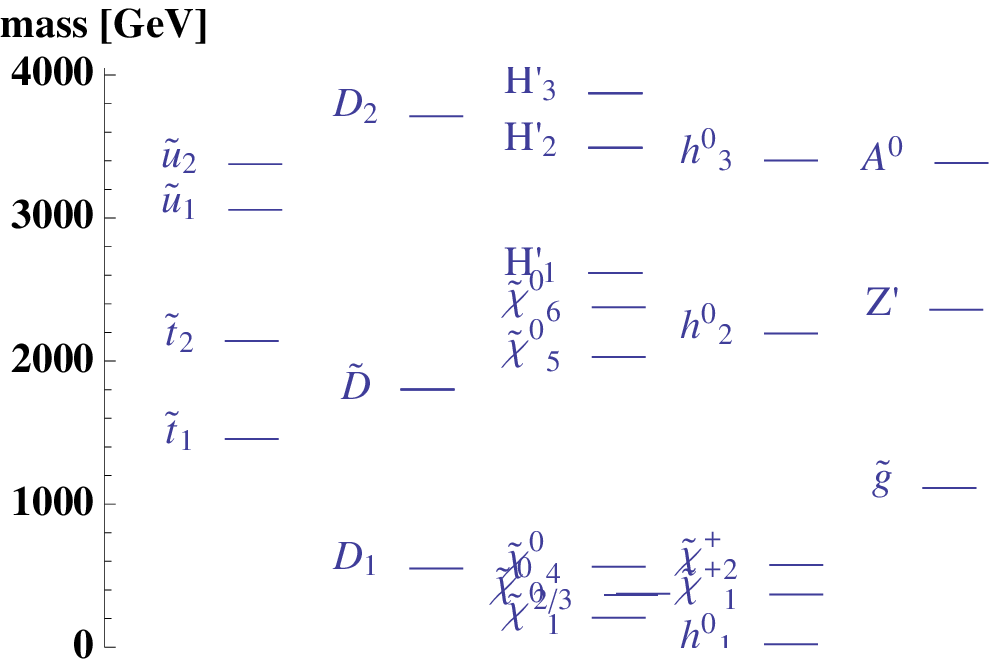}
\hspace{.5cm}
\includegraphics[width=.45\linewidth]{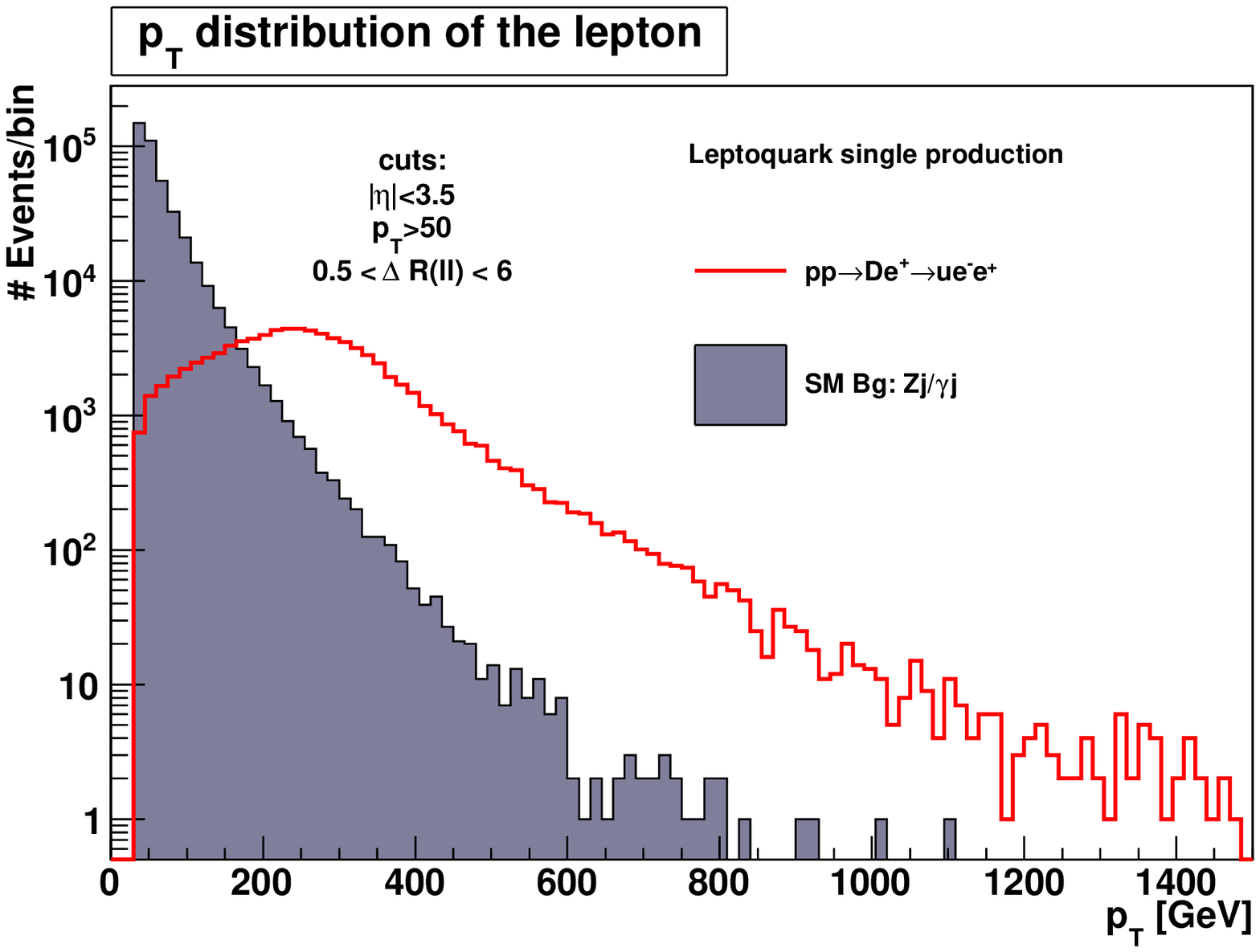}
\caption{\label{tab:spec_pt} 
Left: Spectrum correponding to the input parameters from Tab.~\ref{tab_params}
Right: Signal-to-background analysis of the $p_T$-distribution of
final state leptons from leptoquark single production.}
\end{figure}

Drawing a completely general conclusion from sparse scans through such
a high-dimensional parameter space is difficult, so we shall
concentrate in the following on a sample scenario, obtained by the
MCMC and discuss some phenomenological features in a little more detail.
Fig.~\ref{tab:spec_pt} shows on the left a sample spectrum, for which
the leptoquark states are rather light, order 600 GeV. The lightest Higgs 
has a mass around 110 GeV, whereas the remaining two Higgs particles are
heavier are far in the decoupling limit at 1 TeV or more. The
$Z^\prime$ mass lies at order 2 TeV, such that should also be discovered  
at an early stage of the LHC. But the really spectacular feature are
the light leptoquarks, which yield cross section in the picobarn
range. With an integrated luminosity of 100 fb$^{-1}$, around 100,000
signal events could be collected in the area of parameter space
presented here. This would allow a spectroscopy of the leptoquark
sector, and maybe even a determination of the Yukawa couplings.

We implemented this Pati-Salam symmetric Supersymmetric Standard Model
(PSSSM) into the multi-purpose event generator WHIZARD~\cite{WHIZARD},
which has been designed especially for BSM analyses, and many SUSY
phenomenology projects have been performed~\cite{whiz_susy}. Here, we
followed basically the conventions in~\cite{slhaetc}. As a
prime example for the LHC phenomenology of this model we show here the
$p_T$ distribution of single leptoquark production in that model. The
chosen parameter point yields a spectacular signal-to-background
ratio of 127.

In summary, we analyzed both the high-dimensional parameter space of
the PSSSM and its LHC phenomenology, which might give striking signals
already at an early LHC stage. More phenomenology has been studied
here~\cite{wiesler}.

\begin{theacknowledgments}
The authors are supported by the Ministerium f\"ur Wissenschaft und Kultur
of the state Baden-W\"urttemberg by the program ZO IV and the German
Research Society (DFG) under grant no. Re 2850/1-1 as well as the DFG
graduate school GRK 1102 ``Physics at Hadron Colliders''.  JR would
like to thank to Aspen Center for Physics for their
hospitality. Special thanks go to Frank Deppisch for providing the
basic RGE code. We are also grateful to S.F.King, R. Nevzorov, 
P. Fileviev-Perez, and W. Kilian for valuable remarks and
discussions.
\end{theacknowledgments}



\bibliographystyle{aipprocl} 

\begin{thebibliography}{99}


\bibitem{Kilian:2006hh}
  W.~Kilian and J.~Reuter,
  Phys.\ Lett.\  B {\bf 642}, 81 (2006).

\bibitem{Howl:2007hq}
  R.~Howl and S.~F.~King,
  Phys.\ Lett.\  B {\bf 652}, 331 (2007).

\bibitem{Pati:1974yy}
  J.~C.~Pati and A.~Salam,
  Phys.\ Rev.\  D {\bf 10}, 275 (1974)
  [Erratum-ibid.\  D {\bf 11}, 703 (1975)].

\bibitem{Griest:1990vh}
  K.~Griest and M.~Sher,
  Phys.\ Rev.\  D {\bf 42}, 3834 (1990).

\bibitem{WHIZARD}
  W.~Kilian, T.~Ohl and J.~Reuter,
  arXiv:0708.4233 [hep-ph];
  M.~Moretti, T.~Ohl and J.~Reuter,
  arXiv:hep-ph/0102195;
  J.~Reuter, PhD thesis
  arXiv:hep-th/0212154.


\bibitem{whiz_susy}
  T.~Ohl and J.~Reuter,
  Eur.\ Phys.\ J.\  C {\bf 30}, 525 (2003);
  K.~Hagiwara {\it et al.},
  Phys.\ Rev.\  D {\bf 73}, 055005 (2006);
  W.~Kilian, J.~Reuter and T.~Robens,
  Eur.\ Phys.\ J.\  C {\bf 48}, 389 (2006);
  T.~Robens, J.~Kalinowski, K.~Rolbiecki, W.~Kilian and J.~Reuter,
  Acta Phys.\ Polon.\  B {\bf 39}, 1705 (2008);
  J.~Kalinowski, W.~Kilian, J.~Reuter, T.~Robens and K.~Rolbiecki,
  JHEP {\bf 0810}, 090 (2008);
  J. Reuter, F. Braam, these proceedings, 
  arXiv:0909.3059 [hep-ph].



\bibitem{slhaetc}
  P.~Skands {\it et al.},
  JHEP {\bf 0407}, 036 (2004);
  B.~Allanach {\it et al.},
  Comput.\ Phys.\ Commun.\  {\bf 180}, 8 (2009);
  J.~A.~Aguilar-Saavedra {\it et al.},
  Eur.\ Phys.\ J.\  C {\bf 46}, 43 (2006).

\bibitem{wiesler}
  D. Wiesler, Master Thesis, Freiburg, July 2009.

\end{thebibliography}


%

\end{document}